\documentclass[12pt]{iopart}

\usepackage{graphicx}
\usepackage{dcolumn}
\usepackage{bm}
\usepackage{hyperref}
\hypersetup{colorlinks=true, citecolor=blue, urlcolor=blue, linkcolor=blue}
\usepackage{color}
\usepackage{dsfont}
\usepackage{subfigure}
\usepackage{graphicx}
\usepackage{epsfig}
\usepackage{epstopdf}
\usepackage[english]{babel}
\usepackage[T1]{fontenc}
\usepackage[utf8]{inputenc}
\usepackage[normalem]{ulem}

\newcommand {\la} {\langle}\newcommand {\ra} {\rangle}
\newcommand {\beq} {\begin{eqnarray}}
\newcommand {\eeq} {\end{eqnarray}}
\newcommand {\eeqn} [1] {\label{#1} \end{eqnarray}}
\newcommand {\eol} {\nonumber \\}
\newcommand {\ve} [1] {\mbox{\boldmath $#1$}}

\begin{document}
\title{Perturbative correction to the adiabatic approximation for $(d,p)$ reactions}

\author{L Moschini, N~K Timofeyuk and R~C Johnson}

\address{Department of Physics, Faculty of Engineering and Physical Sciences, University of Surrey, Guildford, Surrey GU2 7XH, United Kingdom}

\ead{l.moschini@surrey.ac.uk}

\begin{abstract}
    The adiabatic distorted wave approximation (ADWA) is widely used by the nuclear community to 
    analyse deuteron stripping ($d$,$p$) 
    experiments.
    It provides a quick way to take into account an important property of the reaction mechanism: deuteron breakup.
     In this work we provide a numerical quantification of a perturbative correction to this theory, recently proposed in [Johnson R C 2014 \textit{J.\ Phys.\ G:Nucl.\ Part.\ Phys.}\ \textbf{41} 094005] for 
     separable rank-one nucleon-proton potentials.
    The correction 
    involves an additional, nonlocal, term in the
    effective 
    deuteron-target
    ADWA potential in the entrance channel.
    %
    We test the calculations with perturbative corrections 
    against continuum-discretized coupled channel predictions which treat deuteron breakup exactly.
    %
\end{abstract}

\section{Introduction}
Nuclear transfer reactions, and   deuteron stripping -- ($d$,$p$) -- in particular,  receive continuous interest by the nuclear community, since they provide an excellent framework to study 
spectroscopy of the nuclei involved in the reaction 
\cite{B1950,BKHN52,J14}. A comprehensive review on the theory of ($d$,$p$) processes can be found in  \cite{TIMOFEYUK2020103738}. One important feature of the $(d,p)$ dynamics is deuteron breakup that happens easily  because of the small deuteron binding energy. Therefore, the entrance-channel deuteron-target  wave function  is often described by the  three-body Watanabe Hamiltonian 
\cite{watanabe}  and an approximate solution of the three-body problem  is used within the
adiabatic distorted wave approximation (ADWA)  \cite{JS1970,JT74}.  The latter is based on the assumption that the deuteron breakup involves transitions to low energy scattering states described by wave functions strongly resembling  the ground state within the small $n$-$p$ separations that give the main contribution to the ($d$,$p$) amplitude. Providing a fast and easy way to obtain the three-body $n$-$p$-target wave function within the small $n$-$p$ range, the ADWA results in cross sections
that typically  differ less than 10\%-20\% from those obtained using more extensive 
three-body solutions, such as obtained in the continuum-discretized coupled channels (CDCC) \cite{RAW1975,AIKKRY87} or Faddeev \cite{Fad,Fad96} approaches that seek three-body solutions spanning all the $n$-$p$ range.
However, there are some  cases  where the ADWA and CDCC results differ  significantly, as shown for instance in \cite{ND2011,UDN2012} and more recently in a systematic study by Chazono, \emph{et al}, \cite{chazono}. For this reason it would be desirable to have a simple and accurate way of correcting the three-body wave function at small $n$-$p$ separations. Such a method would be very useful for getting better quality fast predictions for $(d,p)$ cross sections either at the stage of experiment planning or during the analysis of experiments already performed. If proved successful, the first-order perturbation theory could also be extended to include nonlocal nucleon-target optical potential within three-body dynamics of $(d,p)$ reactions. Currently, there are no CDCC developments that allow this. On the other hand, first-order perturbation theory could also be used to go beyond adiabatic approximation in treating explicit energy-dependence of optical potentials  and induced three-body force as proposed in \cite{Joh14} and \cite{Din19}, respectively.

 
  The aim of the present paper is to correct the adiabatic deuteron-target wave function using perturbation theory ideas, more precisely, those proposed
 in \cite{J14}. The  first-order perturbation formalism described there 
 has never been assessed numerically, so in this work we present 
its first quantitative study.  
In section~\ref{model} we recall the effective potential from \cite{J14} and in section~\ref{numerics} we detail the computational approach applied in our calculation. In section~\ref{results} we discuss perturbative  results for several  ($d$,$p$) processes, comparing them to ADWA and CDCC calculations. 
We summarise our findings in section~\ref{conclusions}. An Appendix provides a derivation of an analytical expression for the Yamaguchi rank-1 s-wave scattering wave functions.


\section{Theoretical background \label{model}}
\subsection{Three-body ($d$,$p$) reaction model and adiabatic approximation}
\begin{figure}[!h]
  \centering
  \includegraphics[scale=0.95]{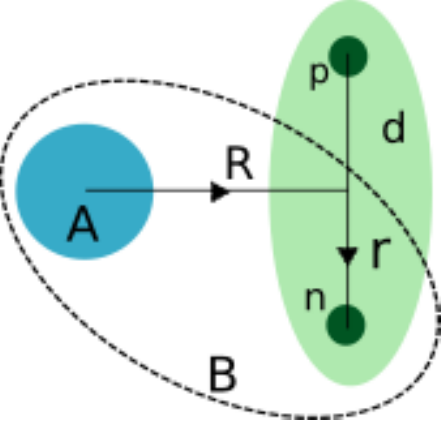}
  \caption{\label{figCoor} Coordinates involved in the $A$($d$,$p$)$B$ reactions: $\mathbf{r}$ is the radius-vector connecting the neutron and proton in the incident deuteron d, and  $\mathbf{R}$ is the radius-vector connecting the deuteron and the target A centres of mass.}
  \end{figure}

In a colliding system as depicted in figure~\ref{figCoor}, 
the $A$($d$,$p$)$B$ scattering amplitude can be written 
\begin{equation}
\label{Tdp}
    T_{dp} = \langle \chi_{\mathbf{k}_p}^{(-)}(p)\phi_p\psi_B(n,A)| V_{np} | \Psi_{\mathbf{k}_d}^{(+)}(n,p) \Psi_A \rangle,
\end{equation}
where  $\chi_{\mathbf{k}_p}^{(-)}(p)\phi_p\psi_B(n,A)$ represents the final state $\psi_B(n,A)$ of nucleus B and an outgoing proton $\phi_p$ with momentum $\mathbf{k}_p$, and $\Psi_{\mathbf{k}_d}^{(+)}(n,p)$ is the  ground state component of the scattering state corresponding to a deuteron with momentum $\mathbf{k}_d$ incident on nucleus $A$ in its ground state $ \Psi_A$. The derivation of this three-body model and the approximations involved are reviewed in \cite{TIMOFEYUK2020103738}.  
This matrix element is dominated by contributions from within the short range of the $n$-$p$ interaction $V_{np}$ and, therefore, these parts of $\Psi_{\mathbf{k}_d}^{(+)}(n,p)$ are emphasised. 
In this case, as shown in \cite{JT74}, it is convenient to expand the $\Psi_{\mathbf{k}_d}^{(+)}(n,p)$ scattering state in a Weinberg basis \cite{weinberg63,weinberg64}
\begin{equation}
   \Psi_{\mathbf{k}_d}^{(+)}(n,p) = \sum_{i=0}^{\infty} \phi^W_i(\mathbf{r})\chi_i^{W(+)}(\mathbf{R}), \label{Wexp} 
\end{equation}
with Weinberg eigenstates $\phi^W_i$ and components  $\chi_i^{W(+)}$ related to $\Psi_{\mathbf{k}_d}^{(+)}(n,p)$ by
\begin{equation}
\label{Wexp2}
  \chi_i^{W(+)}(\mathbf{R})=-\frac{\langle  \phi^W_i|V_{np}|\Psi_{\mathbf{k}_d}^{(+)}(n,p)\rangle}{\langle  \phi^W_i|V_{np}|\phi^W_i\rangle}, 
 \end{equation}
  where in this equation the notation $\langle\dots\rangle $ means integration over $\mathbf{r}$ only. 
  
The validity of the ADWA is based on the accuracy of two observations.
    
(i) A detailed study in  \cite{PTJT13} showed that for many reactions of physical interest the matrix element in (\ref{Tdp}) is dominated by the $i=1$ term in the expansion (\ref{Wexp}). Using the fact that the $i=1$ Weinberg state is just the deuteron ground state $\Psi_d$ and the special orthogonality property of Weinberg states \cite{JT74} we find
\begin{equation}
\label{Wexp3}
  \chi_1^{W(+)}(\mathbf{R})=-\frac{\langle  \Psi_d|V_{np}|\Psi_{\mathbf{k}_d}^{(+)}(n,p)\rangle}{\langle  \Psi_d|V_{np}|\Psi_d\rangle}. 
  \end{equation}
 Consistent with this result the ADWA ignores all explicit contributions to $T_{dp}$ with $i\neq 1$.

(ii) In the ADWA the further approximation is made of ignoring the coupling between different Weinberg components, in which case the  $\chi_1^{W(+)}(\mathbf{R})$ becomes $\chi^{\rm ADWA}(\mathbf{R})$ and is given exactly by the solution of the equation 
\begin{equation}
  \label{schr_adwa_wf}
  \left[T_{R} + V_{\rm ADWA}(\mathbf{R}) -E^+_d \right] \chi^{\rm ADWA}(\mathbf{R}) = 0,
\end{equation}
where $E^+_d$ is the deuteron incident energy in the system c.m.\ frame, $T_{R}$ is the kinetic energy operator associated to $R$, and adiabatic distorted potential is given in \cite{JT74} by
\begin{equation}
    V_{\rm ADWA}(\mathbf{R}) = \frac{\langle\Psi_d |  V_{np}(\mathbf{r}) V_{Ad}(\mathbf{R},\mathbf{r}) |\Psi_d \rangle}{\langle \Psi_d  | V_{np}(\mathbf{r})| \Psi_d \rangle }.
    \label{Vadwa}
\end{equation}
The ADWA potential depends on the deuteron wave function $\Psi_d(\mathbf{r})$, 
 the $n$--$p$ interaction $V_{np}(\mathbf{r})$, and the interaction between the target $A$ and the deuteron components $V_{Ad}(\mathbf{R},\mathbf{r})=V_{nA}(\mathbf{R}+\frac{1}{2}\mathbf{r})+V_{pA}(\mathbf{R}-\frac{1}{2}\mathbf{r})$. In the zero-range limit for $V_{np}$ this reduces to the Johnson-Soper potential $V_{\rm ADWA}(\mathbf{R}) = V_{nA}(\mathbf{R})+V_{pA}(\mathbf{R})$  \cite{JS1970}.
 
 In this paper we study the validity of (i) and (ii) for the special case that $V_{np}$ is the Yamaguchi rank-1 separable interaction  \cite{yamaguchi}
\begin{equation}
V_{np}=-|f_{np}\rangle\langle f_{np} |,
    \label{Vyamaguchi} 
    \end{equation}
where
 \begin{equation}
\langle \mathbf{r} | f_{np} \rangle = N_1 \exp(-\beta r)/r.
    \label{Vyamaguchi2} 
    \end{equation}
 where $N_1$ and $\beta$ are chosen to give the correct deuteron binding energy $\epsilon_d$ and give a good fit to the low energy $n-p$ $s$-wave scattering.

\subsection{First order perturbation to the adiabatic potential for Yamaguchi interaction}
A feature of rank-1 separable potentials is that step (i) is exact because then in $T_{dp}$
\begin{equation}
V_{np}\Psi_{\mathbf{k}_d}^{(+)}(n,p)=-|f_{np}\rangle\langle f_{np} |\Psi_{\mathbf{k}_d}^{(+)}(n,p),
    \label{Vyamaguchi3} 
    \end{equation}
    and the right-hand-side of Eq. (\ref{Wexp3}) reduces to 
 \begin{equation}
 \label{Wexp4}
 -\frac{\langle  \Psi_d|V_{np}|\Psi_{\mathbf{k}_d}^{(+)}(n,p)\rangle}{\langle  \Psi_d|V_{np}|\Psi_d\rangle}=\frac{\langle f_{np}|\Psi_{\mathbf{k}_d}^{(+)}(n,p)\rangle}{\langle f_{np}|\Psi_d\rangle}. 
 \end{equation}
 Hence, Eq. (\ref{Vyamaguchi3}) can be written
 \begin{equation}
V_{np}\Psi_{\mathbf{k}_d}^{(+)}(n,p)=V_{np}|\Psi_d\rangle\chi^\mathrm{eff}(\ve{R}),
    \label{eq11} 
    \end{equation}
    where
  \begin{equation} \chi^\mathrm{eff}(\ve{R})=\frac{\langle f_{np}|\Psi_{\mathbf{k}_d}^{(+)}(n,p)\rangle}{\langle f_{np}|\Psi_d\rangle}.   
  \label{eq12}
  \end{equation}  
 A second feature of the rank-1 model $V_{np}$ is that, as demonstrated in \cite{J14}, an effective deuteron-target distorting potential that generates $\chi^\mathrm{eff}(\ve{R})$ exactly and hence goes beyond $V_{\rm ADWA}$ can be derived. A perturbative approach \cite{J14} to the calculation of this effective distorting potential is described in the next Section.
 
     The choice of $V_{np}$ interaction is found to have a small impact on $(d,p)$ cross sections 
\cite{HKB04,JBF08,GT18,Dd18}, so we follow \cite{J14} and use the Yamaguchi rank-1 separable $n$-$p$ potential with parameters that give the corrrect binding energy of the deuteron, its root-mean-squared radius and the asymptotic normalization constant (ANC). The Yamaguchi ground state wave function is identical to the Hulth\'en deuteron wave function that reads \cite{hulthen1,hulthen2}
\begin{equation}
    \Psi_d(\mathbf{r}) = N_0 \frac{e^{-\alpha r}-e^{-\beta r}}{r},
    \label{gsWF}
\end{equation}
where   $\alpha= \sqrt{2 \mu_d \epsilon_d}/\hbar$ with $\mu_d$ being the reduced mass of the $n$-$p$ system, $|\epsilon_d|=2.2$~MeV,  $\beta = 6.255 \alpha$ and $N_0$ normalizes the deuteron wave function to unity.


The s-wave scattering wave function of energy $E_k= \hbar^2 k^2 /\mu_d$ generated by the Yamaguchi potential is given by (see  \ref{appendix})
\begin{equation}
    \chi_{l=0}(k,r) = \frac{e^{i\delta_k}}{kr} \left[ \sin(kr+\delta_k)-\sin \delta_k e^{-\beta r}  \right],
        \label{scattWF}
\end{equation}
where the phase shift $\delta_k$ is
\begin{equation}
   k \cot(\delta_k) = \frac{k^4 + (\alpha^2 +2\alpha\beta +3\beta^3)k^2 -\alpha \beta^2(\alpha+2\beta)}{2  \beta (\alpha+\beta)^2}.
    \label{cotdelta}
\end{equation}
We have checked that for $n$-$p$ relative energies up to 25~MeV these phase shifts are in excellent agreement (better than 1.5$\%$)  with those obtained in a simplified  $n$-$p$ model,  given by a single Gaussian interaction with the depth of $72.15$~MeV  and the range of $1.484$~fm  \cite{Dgauss},   widely used in the CDCC treatment of $(d,p)$ reactions.

In the perturbative approach of \cite{J14}, to the first order in $\Delta V= V_{Ad}(\mathbf{R},\mathbf{r}) - V_{\rm ADWA}(\mathbf{R})$  the effective potential that generates $\chi^\mathrm{eff}$  is given by
\begin{equation}
    V_{\rm eff}(\mathbf{R}) \approx V_ {\rm ADWA}(\mathbf{R}) + G^{\rm ADWA}(E^+_d -{\bar \epsilon})\delta V,
    \label{Veff}
\end{equation}
where $ V_ {\mathrm ADWA}$ is defined in (\ref{Vadwa}).
The first factor of the correction term is the Green function
\begin{equation}
G^{\rm ADWA}(E^+_d - {\bar \epsilon}) = \frac{1}{E^+_d - {\bar \epsilon} -T_{R} - V_{\rm ADWA}} ,
\end{equation}
where the threshold 
   ${\bar \epsilon} = \langle f_{np}| T_r | \Psi_d\rangle/\langle f_{np}|\Psi_d\rangle
   \label{epsilon}$ 
is the energy in the $n$-$p$ continuum at which the corresponding  $n$-$p$ scattering states first show an oscillatory dependence on r within the range of $V_{np}$ and differ significantly from that of the deuteron ground state wave function for the same range of $\mathbf{r}$ \cite{J14}. 
The other factor of the corrective term in (\ref{Veff}) is defined as
\begin{equation}
\delta V =\frac{\langle f_{np}| T_r \Delta V | \Psi_d\rangle}{\langle f_{np}|\Psi_d\rangle},
\label{deltaV}
\end{equation}
with $T_r$ the kinetic energy operator associated with $r$.
For the Yamaguchi $n$-$p$ potential and  Johnson-Tandy $A$-$d$ potential,  we find
\begin{equation}
\delta V_A(R) = \frac{\hbar^2 \beta^2}{\mu_d} \left(1+\frac{\alpha}{\beta} \right)  \left[ V_{An}(R)+V_{Ap}(R) - V_{\rm ADWA}(R)\right].
\label{deltaV-eq48}
\end{equation}

The effective distorted wave $\chi^{\rm eff}(\mathbf{R})$ is then obtained solving the non-homogeneous integral-differential Schr\"odinger equation 
\begin{equation}
  \label{schr_corr_wf}
  \left[T_{R} +V_C(R)+ V_{\rm ADWA}(R) -E^+_d \right] \chi^{\rm eff}(\mathbf{R}) = -s(\mathbf{R})
\end{equation}
with the source term 
\begin{equation}
 s(\mathbf{R}) = \int G^{\rm ADWA}(\mathbf{R},\mathbf{R}')  \delta V_A(\mathbf{R}') \chi^{\rm eff}(\mathbf{R}') d\mathbf{R}'.
  \label{source-term}
\end{equation}
Equation (\ref{schr_corr_wf}) also involves the deuteron-target Coulomb interaction potential $V_C(R)$. As in all $(d,p)$ calculations we use  the Coulomb potential of a uniformly-charged sphere for $V_C$ with the Coulomb radius of $R_C=1.3 \cdot A^{1/3}$~fm, where $A$ the target mass number.

\section{Numerical aspects of perturbative calculation \label{numerics}}


In this section we  provide important details of the $\chi^{\rm eff}$ calculation. 
The calculated wave function is then read in as an input by computer code  TWOFNR \cite{TWOFNR} and used to evaluate the $(d,p)$ $T$-matrix and the corresponding corrected cross sections. 

 \begin{figure}
     \centering
     \includegraphics[scale=0.45]{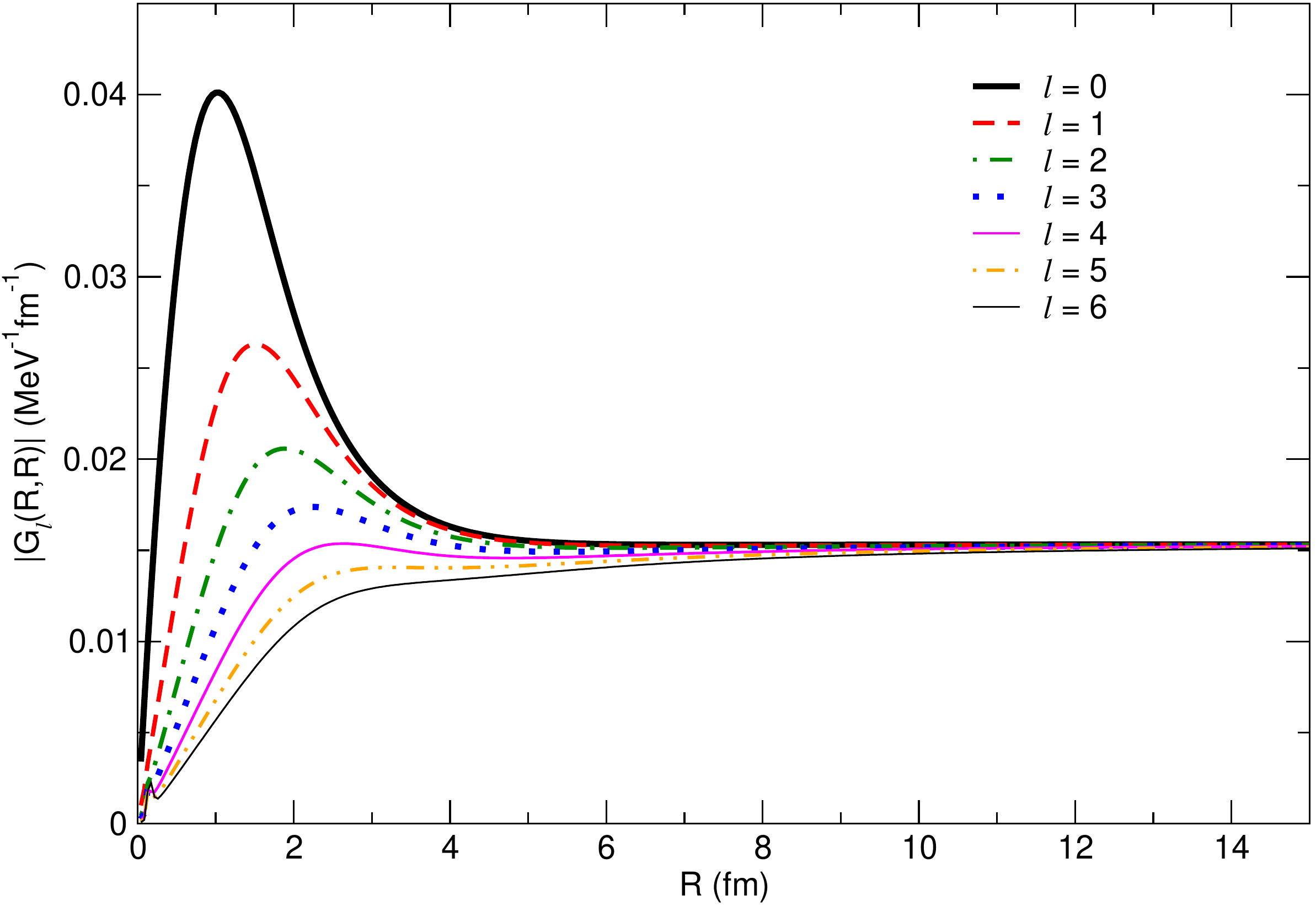}
     \caption{Absolute value of diagonal Green's  function, for partial waves up to $l=6$, calculated for $d+^{10}$Be system at the incident deuteron energy of 40.9 MeV.
     }
     \label{fGRR}
 \end{figure}
 
\subsection{Evaluation of the Green function}
The effective potential $V_{\rm eff}$ 
 involves a Green function that can be calculated using  the partial-waves expansion  \cite{satchler-DNR} 
\begin{equation}
\label{GF}
    G^{\rm ADWA}_l(R,R') = -\frac{2\mu}{\hbar^2 \kappa} \frac{{\cal G}^{(1)}_l(\kappa R_<){\cal G}^{(2)}_l(\kappa R_>)}{{\cal W}_l},
\end{equation}
where $\mu$ is the $A$-$d$ reduced mass, $\kappa = \sqrt{2\mu(E^+_d - {\bar \epsilon})/\hbar^2}$,  and ${\cal G}^{(1)}_l$ and ${\cal G}^{(2)}_l$ are linearly independent solutions of the  equation 
\beq
    \left[ T_l + V_C(R) + V_{\rm  ADWA}(R) - (E^+_d - {\bar \epsilon}) \right]{\cal G}_{l}(R) = 0
\eeqn{eqGl}
with the kinetic energy operator in the partial wave $l$,
\beq
T_l =-\frac{\hbar^2}{2\mu}\frac{d^2}{dR^2} + \frac{\hbar^2}{2\mu}\frac{l(l+1)}{R^2}.
\eeqn{Tl}
The quantity ${\cal W}_l$ in (\ref{GF}) represents the  Wronskian of ${\cal G}^{(1)}_l$ and ${\cal G}^{(2)}_l$. 
For the Yamaguchi potential considered in this work the threshold energy, calculated in \cite{TJ13} using the Hulth\'en wave function, is ${\bar \epsilon} = 114$~MeV. Note that this value is strongly model dependent  \cite{BTT2016}: for other choices of the $n$-$p$ interaction it would be different but the perturbative formalism would change, too.
Since most modern experiments are performed at
$E^+_d < {\bar \epsilon}$ we will restrict ourselves by considering such cases only.
The corresponding  functions ${\cal G}^{(1,2)}_l$ 
satisfy boundary conditions  ${\cal G}^{(1)}_l(R) \sim  R^{l+1} $ near $R \approx 0$ and ${\cal G}^{(2)}_l(R)  = W_{-\eta,l+1/2}(2 \kappa R)$ as $R \rightarrow \infty$. Here,
 $W$ is
the Whittaker function \cite{AS70} with Sommerfeld parameter $\eta =Ze^2/(4\pi \epsilon_0 \hbar v)$, $Ze$ target charge and $v$ incident velocity, so that the Green function decays exponentially at $R$ outside the range of  $\Delta V$ and $V_{\rm ADWA}$. We have calculated ${\cal G}^{(1)}_l$ and  ${\cal G}^{(2)}_l$  from (\ref{eqGl}) using  finite difference method and 
 we checked that the diagonal Green function  convergence to the same constant value $ |G^{\rm ADWA}_l(R,R)| \rightarrow \mu/(\hbar^2 \kappa)$ for all partial waves at large $R$, as shown in figure~\ref{fGRR}.

Before going further we should note that for large deuteron incident energies, when $E^+_d > \bar{ \epsilon}$, the  ${\cal G}^{(2)}_l$ has a different boundary condition, ${\cal G}^{(2)}_{l}(R) \sim  F_{l}(\kappa R)-iG_{l}(\kappa R)$,
 defined by the regular and irregular Coulomb functions $F_{l}$ and $G_{l}$ respectively \cite{AS70}. This Green's function, and the corresponding perturbative correction to the effective potential, behaves asymptotically like an oscillating outgoing wave with wave number $\kappa$.

\subsection{Evaluation of the distorted wave \label{sec:chi}}

To determine the effective distorted waves $\chi^{\rm eff}_l $, needed in the calculation of the ($d$,$p$) amplitude, we solve  (\ref{schr_corr_wf}) with an iterative method for each partial wave $l$, as suggested e.g.\ in  \cite{M09}. We assume that
the radial distorted wave in the partial-wave expansion of $\chi^{\rm eff}(\mathbf{R})$ is given by $\chi^{\rm eff}_l(R) \equiv u^{\rm eff}_l(R)/R= u^{(n \rightarrow \infty)}_l(R)/R$ and $u^{(n)}_l$ is found from  solution of the iterative problem
\begin{eqnarray}
  \label{schr_corr_wf2}
  \left[
  T_l+V_C(R) + V_{\rm ADWA}(R) -E_d \right] u^{(n+1)}_l(R) = \nonumber \\ 
  -\int_0^{+\infty} G^{\rm ADWA}_l(R,R')  \delta V_A(R') u^{(n)}_l(R') dR'.
  \label{chieffl}
\end{eqnarray}
The first iteration of this equation solves the homogemeous ADWA equation for $u^{(1)}_l$ by setting $u^{(0)}_l$ in the r.h.s.\  of (\ref{chieffl})  to zero. Then using  $u^{(1)}_l$ in the r.h.s.\ we obtain  $u^{(2)}_l$. Repeating this process several times we arrive at a converged solution  for $u^{\rm eff}_l$. The convergence is obtained in less than $10$ iterations, with an accuracy of the first one being more than 95\%. The typical computational time is a couple of seconds, when $40$ partial waves are included. In these calculations we used finite-difference Numerov or Runge-Kutta methods and
imposed the boundary conditions of $u^{(n)}_l(0) = 0$ and  $u^{(n+1)}_l(R_{\max}) = u^{(n)}_l(R_{\max})$. We determined  the $S$-matrix elements $S_l$ from  matching  the normalization of the distorted wave $u^{(n)}_l$ at large distances to $\frac{i}{2} \left( H_l^* - S_lH_l \right)$,
where $H_l = G_l + iF_l$ is a combination of regular and irregular Coulomb functions \cite{satchler-DNR, AS70}.


\begin{figure}[!t]
  \centering
  \includegraphics[scale=0.54]{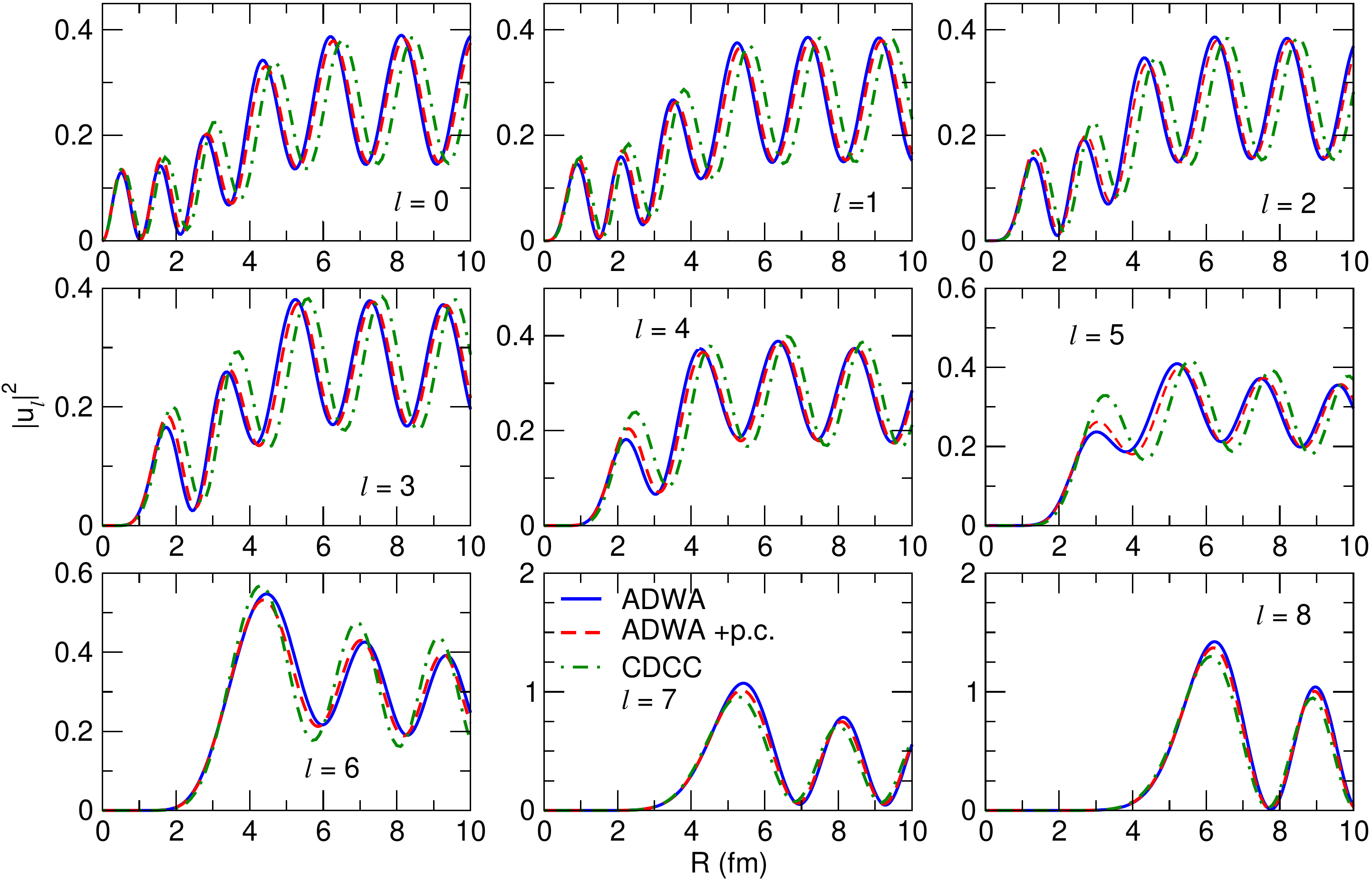}
  \caption{\label{chi}
  ADWA distorted waves (solid lines) compared to the distorted waves $u^{\rm eff}_l$ obtained with perturbative correction (p.c.)  (dashed lines)
  for  $d$+$^{10}$Be scattering at
  $E_d=40.9$~MeV
 shown for partial waves $\ell \leq 8$. }
  \end{figure}

To check the quality of the solution for $u^{\rm eff}_l$ we used an equivalent way of solving (\ref{schr_corr_wf2}), which  is provided  by the system of the coupled equations
\begin{eqnarray}
    \left[ 
    T_l+V_C +V_{\rm ADWA} -E^+_d \right] u_l^{\rm eff}(R) = C_l u_l^{(\rm a)}(R) \\ 
    \left[ 
    T_l+V_C +V_{ADWA} -(E^+_d -\bar{\epsilon}) \right] u_l^{(\rm a)}(R) 
      = C_l^{-1} \delta V(R) u_l^{\rm eff}(R), \nonumber
    \label{cc}
\end{eqnarray}
where $u_l^{(\rm a)}$ is an auxiliary function with the boundary conditions of $u_l^{(\rm a)}(0) =0$ and $u_l^{(\rm a)}(R_{\max}) \sim W_{-\eta,l+1/2}(2\kappa R_{\max})$ while $C_l$ is an arbitrary constant (fixed here to $1$~MeV). 
This system could be derived by introducing the function $u_l^{(\rm a)} =  -C_l^{-1} s_l $, with $s_l$ being the radial part of the source term defined in (\ref{source-term}), and making use of the property of the Green's functions
\begin{equation}
    \left[
    T_l-V_C(R) - V_{\rm ADWA}(R) -(E^+_d - {\bar \epsilon}) \right] G^{\rm ADWA}_l(R,R') = \delta(R,R') .
  \label{GFdelta}
\end{equation}
We solved  these coupled equations using the $R$-matrix code developed by P.~Descouvemont \cite{Rmatrix}. The resulting functions $u^{\rm eff}_l$ were in excellent agreement with those obtained by iterative solution of (\ref{schr_corr_wf2}). An example of $u^{\rm eff}_l$ is shown in figure \ref{chi} for the  $d+^{10}$Be system at $E_d = 40.9$ MeV for a few partial waves, up to $\ell=8$, using global parameterisation CH89 \cite{VTMLC91} for $n$-$^{10}$Be and $p$-$^{10}$Be optical potentials.

Figure  \ref{chi} shows that the character of the distorted waves changes after $\ell=6$, which corresponds to the product $kR_N$ of the deuteron momentum $k \sim 2$ fm$^{-1}$ and nuclear radius $R_N \sim 3$ fm. Distorted waves with  $\ell < 6$ have noticeable presence in the nuclear interior and their imaginary parts significantly contribute to the absolute values $|u_l|^2$. At  $\ell > 6$   all the distorted waves are practically zero within the range $0 < R < R_N$ and they also become real. We also notice that perturbative correction for $\ell > k R_N$ is practically negligible in the internal region. On close examination, we find that absolute value of  $|u_l|^2$ in their maxima are affected by perturbative correction by less then 5$\%$, 2$\%$ and 1$\%$ at $\ell=7$, 8 and 10. The influence on S-matrix element is the largest at $\ell \sim k R_N$, around 8\%,  for  partial waves with $\ell < k R_N$ it does not exceed 5\% and quickly becomes negligible with increasing $\ell$. The S-matrix elements are shown in Table \ref{tab:S}.
To understand this result we go back to (\ref{schr_corr_wf}) with the source term (\ref{source-term}). The integrand of the source term contains the product of the correction potential $\delta V_A$, which is concentrated on the nuclear surface $R_N$,  and the distorted wave that starting from some $\ell$ is zero everywhere within $R_N$. Therefore, for $l > kR_N$ the source term becomes close to zero leading to negligible influence from the perturbative corrections. We have observed the same trend in all other cases studied in this work. This observation  should have certain consequences for the $(d,p)$ cross sections calculations. We can expect that the cross sections calculated with $\ell > kR_N$ will not be affected by the perturbative corrections while those  that retain $\ell \leq k R_N$ only will show maximum sensitivity to them. The final conclusion about the role of perturbative corrections will therefore depend on relative contribution of the lowest partial waves to the $(d,p)$ cross section.  
Below we check this conjecture by using $u^{\rm eff}_l$  to calculate $(d,p)$ cross sections for different ranges of $\ell$ for a variety of targets and deuteron incident energies.\\

\begin{table}[t]
    \centering
    \begin{tabular}{cl|cccccc}
    \hline
     $l$ & & 0& 1& 2& 3& 4& 5\\ \hline
    $|S_l| $ & ${\rm ADWA}$ & 0.2421& 0.2320& 0.2229& 0.1988& 0.1828& 0.1488\\
         &${\rm ADWA+p.c.}$ & 0.2298& 0.2219& 0.2149& 0.1971& 0.1879& 0.1554\\
\hline
  $l$ & & 6& 7& 8& 9& 10 & 11\\ \hline
    $|S_l| $ & ${\rm ADWA}$  &  0.1867& 0.6059& 0.8369& 0.9320& 0.9709 &0.9874\\
         &${\rm ADWA+p.c.}$  &  0.1917& 0.5594& 0.7980& 0.9087& 0.9580 & 0.9805\\
       \hline
    \end{tabular}
    \caption{Absolute value of the $S$-matrix element $|S_l|$ in partial wave $\ell$ calculated for $d+^{10}$Be scattering at $E_d = 40.9$ MeV in ADWA without and with perturbatice correction (p.c.).}
    \label{tab:S}
\end{table}


\section{Perturbative calculations versus ADWA and CDCC \label{results}}
All $(d,p)$ calculations presented in this section were performed using global nucleon optical potentials in the entrance and exit channels, given by either the CH89   \cite{VTMLC91} or KD02 \cite{KD02} and neglecting spin-orbit interaction in the deuteron channel. To describe the bound state wave function of the transferred neutron we used a two-body Woods-Saxon central 
potential model of a standard geometry (radius $r_0 = 1.25$~fm and diffuseness $a = 0.65$~fm), adjusting its depths to reproduce the neutron separation energy of the relevant state. The spectroscopic factor was assumed to be one everywhere.
We compare the $(d,p)$ angular distributions obtained in perturbative model to those given by uncorrected ADWA theory with the Johnson-Tandy potential. To understand  if perturbative correction treatment of $(d,p)$ reactions is sufficient to include significant part of the $n$-$p$ continuum effects, we compare the corresponding cross sections 
to the CDCC predictions made with the help of the computer code FRESCO \cite{FRESCO}. We point out that the rank-1 Yamaguchi $n$-$p$ potential is nonlocal while FRESCO has been developed for local potentials only. Therefore, additional efforts are needed   to perform  such calculations.  We will describe important details of the CDCC calculations  before showing any numerical results.

\subsection{Details of the CDCC calculations}

The CDCC scattering wave function has the structure
 \begin{equation}
   \Psi^{\mathrm{CDCC}(+)}_{\mathbf{k}_d}(n,p) = \sum_{b} \phi_b(r)\chi_b(\mathbf{R})+\dots, \label{CDCC1}
\end{equation}
where the sum is over a set of discretized bound and continuum s-wave eigenstates of $V_{np}$, and the dots indicate non-s-wave continuum states that may contribute to the CDCC coupled equations that determine the $\chi_b(\mathbf{R})$ but do not contribute to $T_{dp}$ because of the s-wave nature of $|f_{np}\rangle.$ 
The functions $\chi_b(\mathbf{R})$ are obtained as solutions of coupled differential equations with
 the coupling matrix elements $V^{\lambda}_{ii'}(R) = \la \phi_{i'} || Y^{*}_{\lambda} (V_{nA}+V_{pA}) || \phi_i\ra$, where $\phi_i$ is either a deuteron ground state wave function or a continuum bin  with orbital momentum $l_i$.
 We calculated these matrix elements externally and then read them into FRESCO. In these calculations the deuteron ground state wave function was taken from the Hulth\'en model while $s$-wave continuum bins $\phi_i$ were constructed using Yamaguchi wave functions given by (\ref{scattWF}). For other $n$-$p$ orbital momenta $l_i$ we used spherical Bessel functions $j_l(kr)$, representing plane waves, to obtain continuum bins.

\begin{figure}
    \centering
    \includegraphics[scale=0.45]{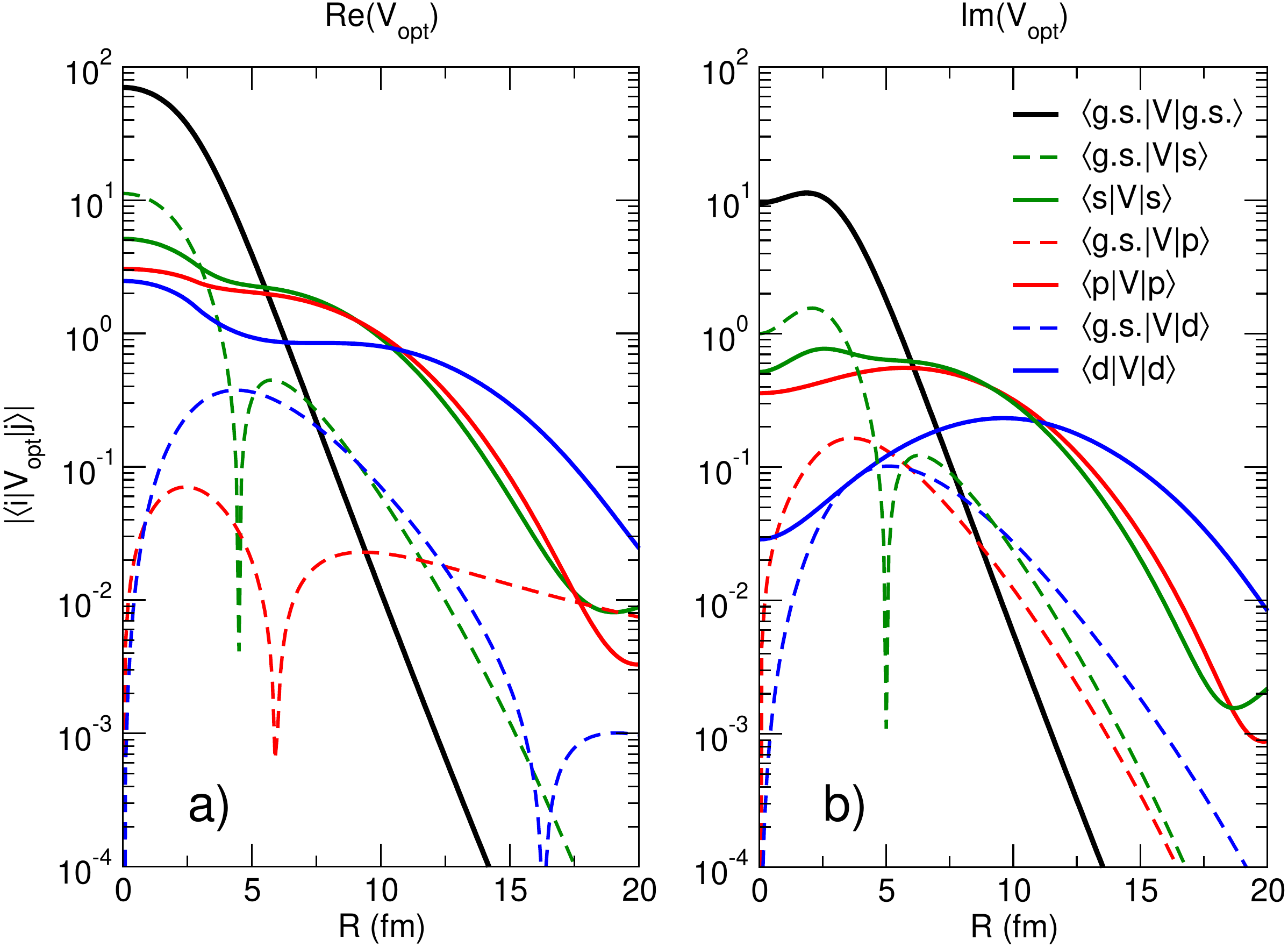}
    \caption{The absolute value of the monopole parts of the coupling potentials $V^{0}_{ii}(R)$ and the coupling potentials connecting deuteron ground state with the lowest  $s$-, $p$- and $d$-wave  continuum bins in the Yamaguchi separable rank-1 $n$-$p$ model.  The calculations are shown for real $(a$) and imaginary $(b)$ parts of the optical potentials separately.}
    \label{fig:coup}
\end{figure}

We show the coupling potentials $V^{\lambda}_{ii'}(R)$ for the lowest $s$-, $p$- and $d$-waves in figure~\ref{fig:coup}, for the case of the $d + ^{10}$Be scattering at $E_d = 40.9$ MeV using nucleon optical potentials $V_{An}$ and $V_{Ap}$  from the Woods-Saxon global optical potential parameterisation CH89  \cite{VTMLC91}. The  $n$-$p$ continuum bins span the energy interval between 0 and 1~MeV. The figure shows that the diagonal ground state $s$-wave matrix element is dominant in the nuclear interior and the coupling between the deuteron ground state and the $p$- and $d$-wave continuum bins is between three and two orders of magnitude smaller in this area. Also, the coupling involving the first $s$-wave continuum bin are two or three time bigger than those involving $p$- and $d$-wave. This suggests that including $p$- and $d$-waves into the coupling scheme will not be important. We have confirmed this later by running the CDCC calculations with $s$-waves only and with including $p$- and $d$-wave continuum. As expected,  the results of such two sets of calculations did not differ.

Unlike in calculations with local $n$-$p$ potentials, where transitions occur explicitly between all continuum bins, the contribution of the CDCC wave function to the  amplitude $T_{dp}$ is determined, from (\ref{eq11}) and (\ref{eq12}), by the quantity
\begin{equation}
V_{np}\Psi_{\mathbf{k}_d}^{(+)}(n,p)=(V_{np}|\Psi_d\rangle)\sum_b\frac{\langle f_{np} | \phi_b\rangle}{\langle f_{np}\mid\Psi_d\rangle}\chi_b(\mathbf{R}).
    \label{Vyamaguchi3CDCC} \end{equation} 
The sum over $b$ represents exactly the expansion over the CDCC basis of the first Weinberg component,  as discussed first in \cite{PTJT13}. The calculations with the first component only are equivalent to those involving well-known one-channel distorted-wave Born approximation.
 We
have  calculated the weight factors (or expansion coefficients) $\frac{\langle f_{np} | \phi_b\rangle}{\langle f_{np}\mid\Psi_d\rangle}$ using the s-wave scattering states of the Yamaguchi potential that appear in the $\phi_b$ and then, using the FRESCO output for $\chi_b(\mathbf{R})$,  carried out the summation  over $b$ to obtain the first Weinberg component. Then, similar to \cite{PTJT13}, this component was read back into FRESCO. The zero-range option for evaluating the $(d,p)$ amplitude was chosen with the standard value $D_0 = -122.50$~MeV~fm$^{3/2}$. 


\subsection{$(d,p)$ cross sections results}
We have tested the perturbative correction on a variety of $(d,p)$ reactions, selecting different targets and different separation energies and quantum numbers of transferred neutron. We have chosen beam energies in the intermediate range, to avoid necessity for including contributions from closed channels. Therefore, all the calculations presented below were carried out at deuteron energies above 40 MeV.
Details on the separation energies $S_n$
and quantum numbers for each system, as well as the corresponding beam
energy and the global parametrisation used are presented in table \ref{tabParam}.
\begin{table}[h!]
    \centering
    \begin{tabular}{cccccc}
 Target &  $S_n$ (MeV) & nlj & $E_d$ (MeV) &  Global param.\\
    \hline
    $^{10}$Be & 0.504 & $2s_{1/2}$ &40.9; 71& CH89\\
    $^{48}$Ca& 5.146 & $2p_{3/2}$ &56; 100& CH89\\
    $^{40}$Ca& 8.363 & $1f_{7/2}$ &56 & KD02\\
    $^{55}$Ni& 16.643 & $1f_{7/2}$ &40 & KD02\\
    $^{40}$Ca \cite{chazono} & 0.1 & $2s_{1/2}$ &40 & KD02\\
  \hline
    \end{tabular}
    \caption{Parameters of the n-target systems used in the calculations. $S_n$ is the
neutron separation energy in MeV,  nlj are the quantum number describing the ground
state of the nucleus in the outgoing channel. The targets are in their $0^+$ ground state, except for $^{55}$Ni, which has a $7/2^-$ ground state. $E_d$ are the deuteron beam energies in MeV considered in each case. We also specify the global parameterisation used in each calculation.}
    \label{tabParam}
\end{table}

\begin{figure}[t]
  \centering
  \includegraphics[scale=0.45]{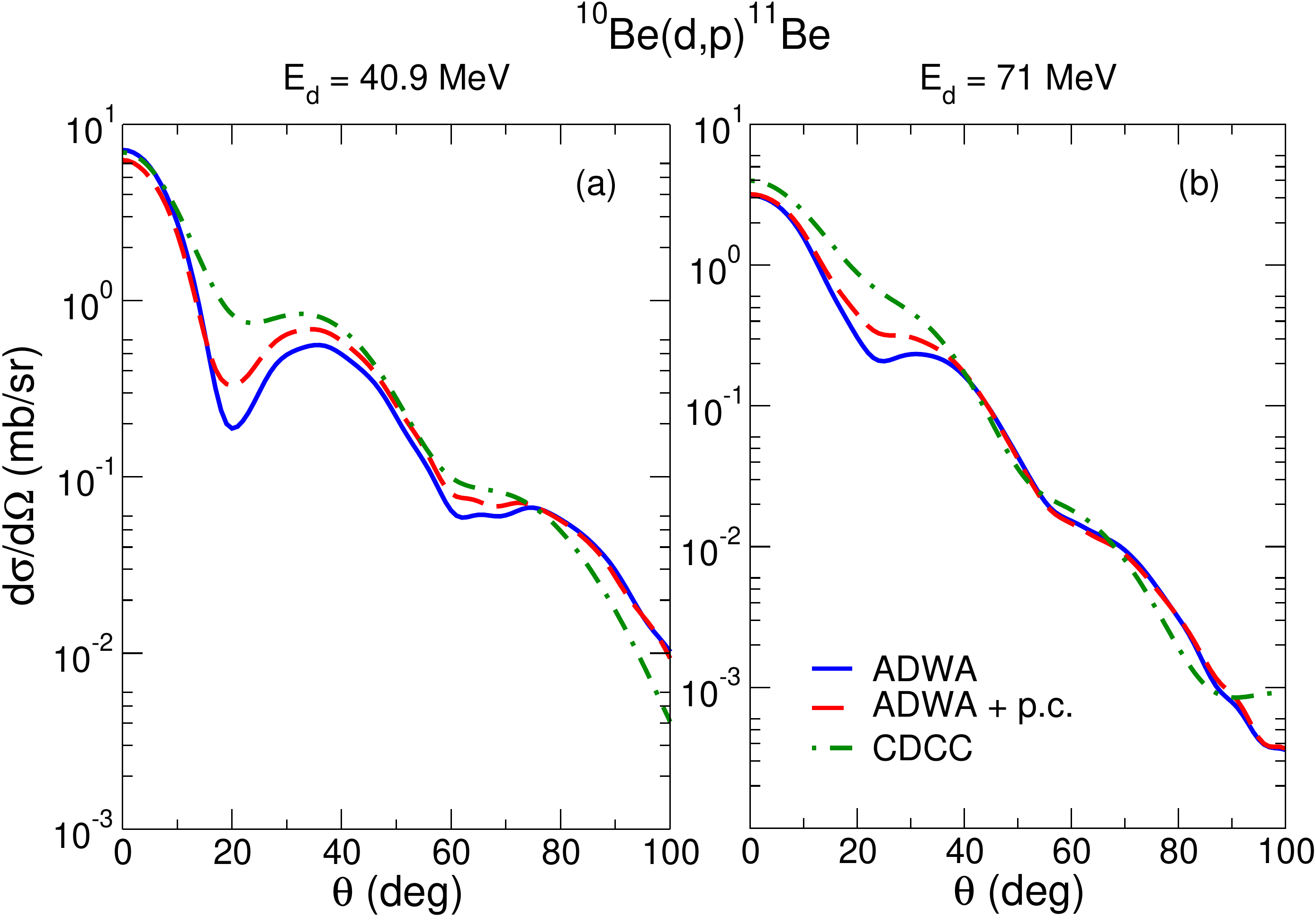}
  \caption{\label{fig10Be} $^{10}$Be($d$,$p$)$^{11}$Be differential cross section with respect to the c.m.\ angle for the calculations with and without the perturbative correction (dashed and solid lines respectively) for $E^{lab}_d=40.9$ and $71$~MeV. The CDCC calculations are shown as  dot-dashed lines. Global parameterisation CH89 was used in these calculations. }
  \end{figure}

\begin{figure}[!h]
  \centering
  \includegraphics[scale=0.45]{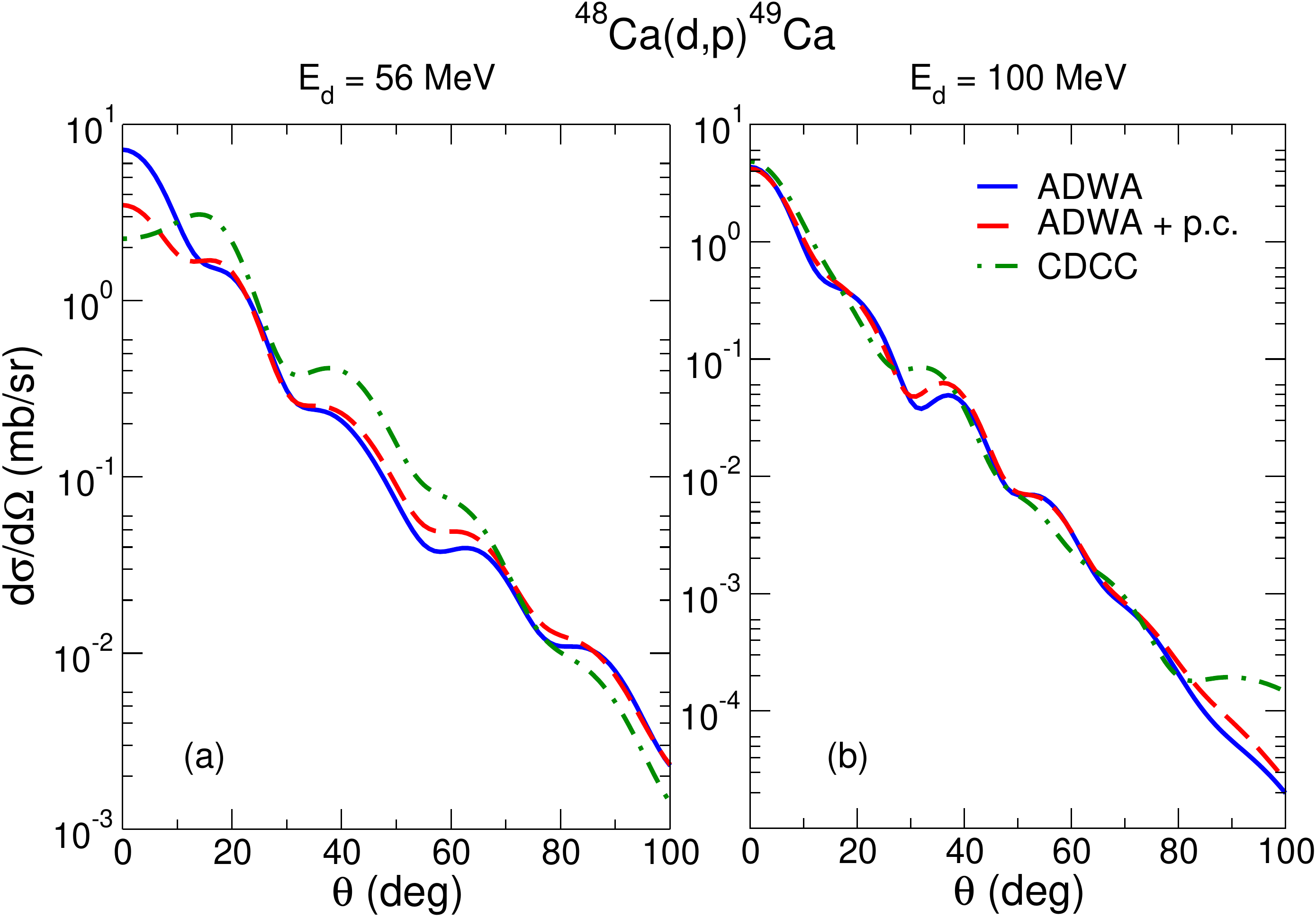}
  \caption{\label{fig48Ca} $^{48}$Ca($d$,$p$)$^{49}$Ca differential cross section with respect to the c.m.\ angle for the calculations with and without the perturbative correction  for $E^{lab}_d=56$ and $100$~MeV (dashed and 
  solid lines respectively). CDCC calculations are shown by dotted-dashed lines. Global parameterisation CH89 was used in these calculations.}
  \end{figure}

\begin{figure}[t]
  \begin{center}
  \includegraphics[scale=0.45]{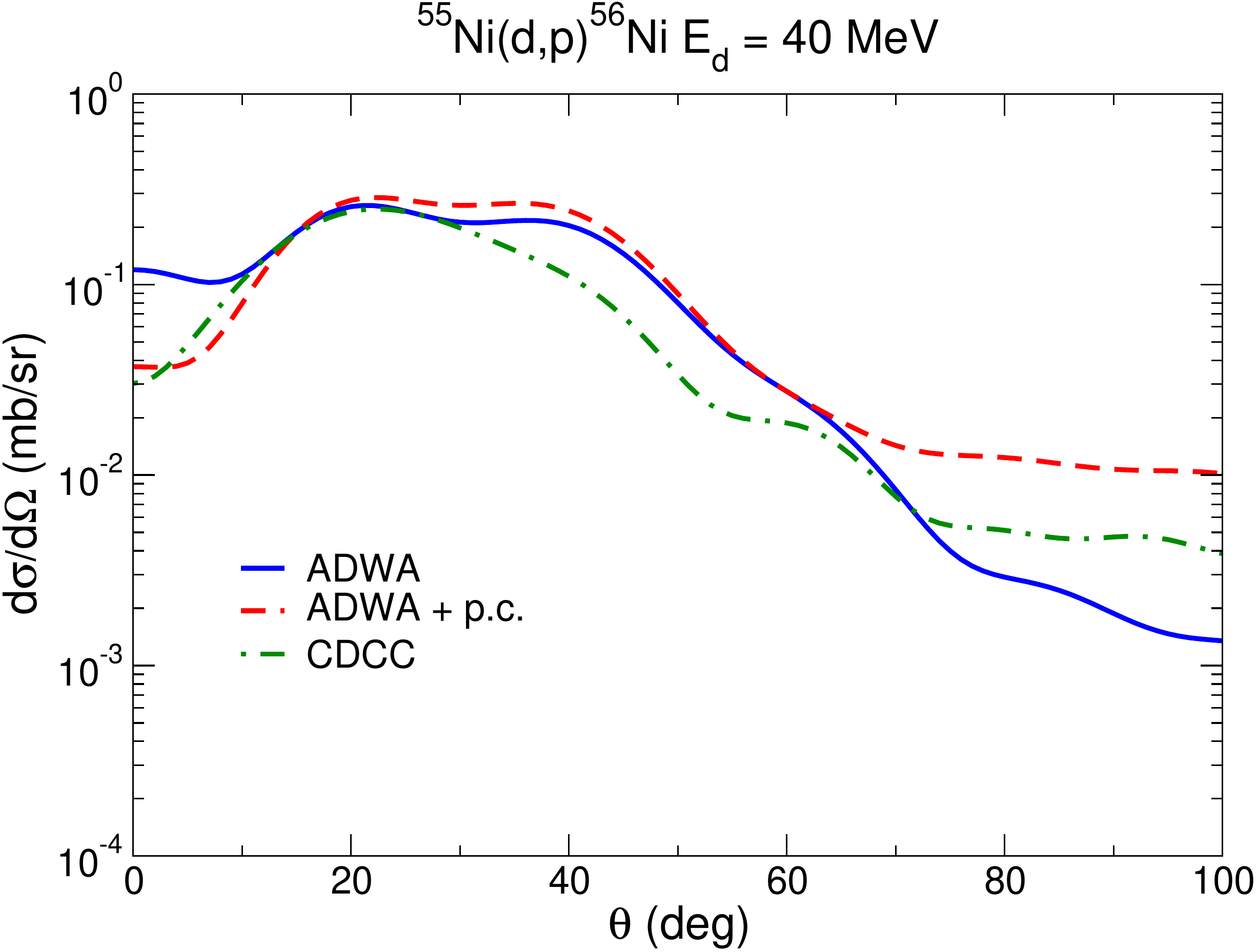}
  \caption{\label{fig56Ni} $^{55}$Ni($d$,$p$)$^{56}$Ni differential cross section with respect to the c.m.\ angle for the calculations with and without the perturbative correction  for $E^{lab}_d=40$~MeV (dashed and 
  solid lines respectively). CDCC calculations are shown by dotted-dashed lines. The global parameterisation  KD02 was used for nucleon optical potentials in the entrance and exit channels. }
    \end{center}
  \end{figure}

\begin{figure}[t]
  \begin{center}
  \includegraphics[scale=0.45]{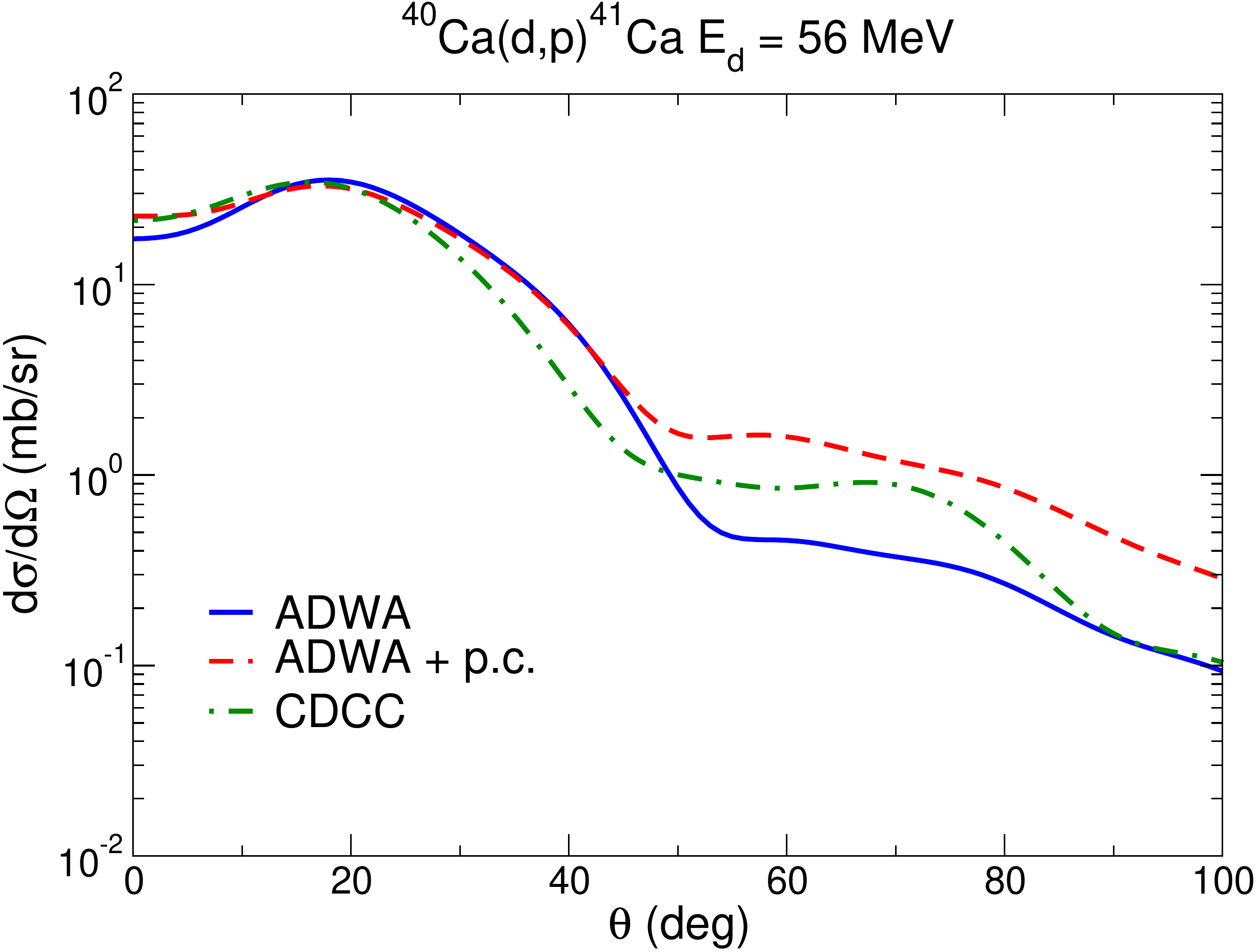}
  \caption{\label{fig40Ca} $^{40}$Ca($d$,$p$)$^{41}$Ca differential cross section with respect to the c.m.\ angle for the calculations with and without the perturbative correction  for $E^{lab}_d=56$~MeV (dashed and 
  solid lines respectively). CDCC calculations are shown by dotted-dashed lines. The global parameterisation  KD02 was used for nucleon optical potentials in the entrance and exit channels. }
    \end{center}
  \end{figure}

We start examining the  $(d,p)$ cross sections results with the $^{10}$Be($d,p)^{11}$Be reaction at $E_d = 40.9$ and $71$~MeV.
In the first case, the distorted waves have already been
discussed in section \ref{sec:chi}. Our calculations with different ranges of partial waves $\ell$, included in description of the deuteron channel, have confirmed the conjecture made in that section that the influence of the perturbative correction on the cross sections comes from the partial waves with $\ell \leq \kappa R_N$ only. For a particular choice of the $^{10}$Be($d,p)^{11}$Be reaction the contribution from the low partial waves is noticeable, with the main contribution coming from $\ell \sim \kappa R_N$. 
The effect of the perturbation on the cross section is compared to the ADWA result in figure \ref{fig10Be}, for both beam energies, as dashed and solid lines respectively. The benchmark CDCC calculations are shown as dot-dashed lines.
In both cases, including perturbative corrections affects mainly the second maximum in a region between 20-50~degrees, with {20-30}$\%$ increase of the cross sections.
  We proceed with the $^{48}$Ca$(d,p)^{49}$Ca reaction, studied for $E_d =56$ and 100~MeV.  Figure \ref{fig48Ca} compares the ADWA, calculated without and with perturbative corrections, with the CDCC predictions. Similar to the $^{10}$Be$(d,p)^{11}$Be case, the perturbative description is not sufficient to get closer to the CDCC results in the angular ranges  where non-adiabatic effects are not negligible.
  The perturbation brings the ADWA result closer to the CDCC at forward angles in the cases of $^{55}$Ni$(d,p)^{56}$Ni reaction at 40~MeV, in figure \ref{fig56Ni}, and $^{40}$Ca$(d,p)^{41}$Ca reaction at 56~MeV, shown in figure \ref{fig40Ca}. However, it is not sufficient at higher angles.
  
 In most of the cases considered above, ADWA and CDCC calculations differ considerably, and thus could be considered non-adiabatic, as already shown e.g.\ by  \cite{ND2011,UDN2012}. While at forward peak the non-adiabatic  effects are of the order of {30}$\%$ in most of the cases, at the second peak they almost double the cross section. In these situations, the effect of the perturbative correction on the second peak is not sufficient to account for this change. By comparing perturbative ADWA and CDCC calculations performed in different fixed ranges of $\ell$ we conclude that perturbative corrections are not sufficient at small $\ell$ as well.
 
\begin{figure}[t]
  \begin{center}
  \includegraphics[scale=0.45]{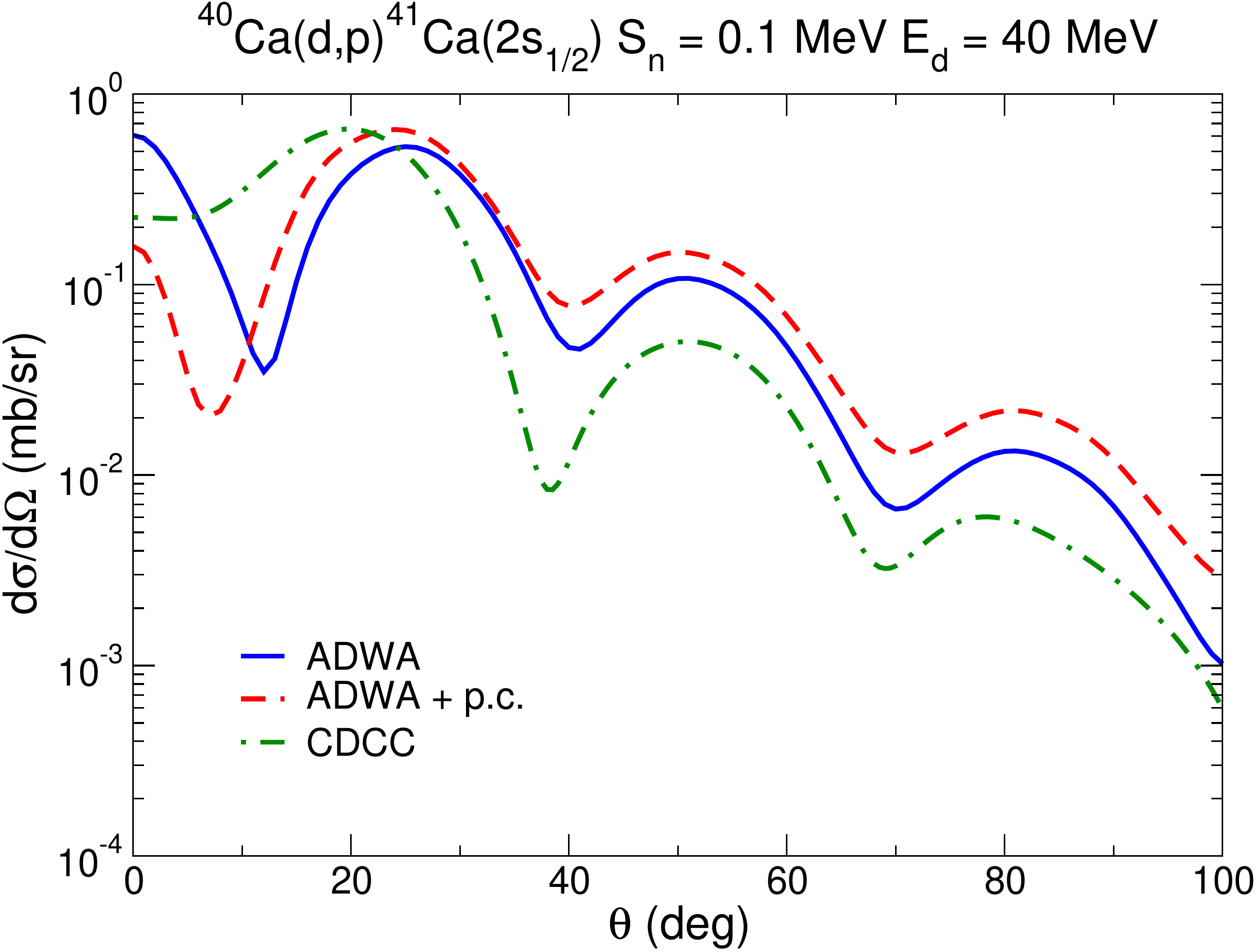}
  \caption{\label{figC} Differential cross section with respect to the c.m.\ scattering angle for a case selected from  systematical study of Chazono et al \cite{chazono}. Solid lines represent the ADWA result while the CDCC calculation is shown as  dot-dashed lines. The result including perturbative correction to the ADWA is shown by dashed lines. The global parameterisation  KD02 was used for nucleon optical potentials in the entrance and exit channels. }
    \end{center}
  \end{figure}

    Other cases where the ADWA and CDCC descriptions differ stronger were identified in  \cite{chazono}, where ADWA validity was  tested systematically  for a variety of cases with different masses, charges, binding energies, angular momenta and beam energies. Among the typical situations in which the ADWA failed significantly highlighted in this study, the one with small $S_n$ and relatively high incident energies was indicated. This was represented by the fictitious  $^{40}$Ca($d$,$p$)$^{41}$Ca(2$s_{1/2}$) with $E_d=40$~MeV and $S_n=0.1$~MeV reaction (figure 5a of \cite{chazono}). We have reproduced the ADWA calculation from \cite{chazono} and compared it with the perturbative calculation and a CDCC benchmark calculation in figure \ref{figC}. The perturbative correction plays a similar role, as the cases considered above. However, the perturbative calculation still differs from the CDCC.

  
  
\begin{table}[h!]
    \centering
    \begin{tabular}{|cc|c|c|c|c|c|c|}
    \hline
    & & \multicolumn{2}{c|}{ADWA+p.c./ADWA}& \multicolumn{2}{c|}{ADWA+p.c./CDCC}& \multicolumn{2}{c|}{ADWA/CDCC}\\
    \cline{3-8}
    Target & $E_d$ (MeV)  & 1st peak & 2nd peak & 1st peak & 2nd peak & 1st peak & 2nd peak\\
    \hline
 $^{10}$Be& 40.9& 0.876& 1.228& 0.905& 0.818& 1.033& 0.666\\
          & 71& 1.018& 1.341& 0.802& 0.697& 0.788&  0.520\\
    \hline
  $^{48}$Ca& 56& 0.485& 1.088& 1.552& 0.568& 3.200& 0.522\\
          & 100 & 0.965& 1.297& 0.868& 0.886& 0.899& 0.683\\
  \hline
$^{55}$Ni& 40& 0.310& 1.101& 1.225& 1.151& 3.956 & 1.045\\
 \hline 
 $^{40}$Ca& 40& 0.261 & 1.230& 0.705& 0.995 &2.703 &0.809 \\
      & 56& 1.314& 0.931&1.054 & 0.956 & 0.802 & 1.027\\   
          \hline
    \end{tabular}
    \caption{The target nucleus, deuteron incident energy (in MeV) and ratios of the corrected ADWA cross section to those calculated in ADWA and CDCC in the first forward peak at 0~degrees and in the second peak at $\theta > 25$~deg. The ratios of the ADWA to CDCC cross sections in the two peaks are given in the last two columns. Note that in the case of $^{40}$Ca target and $E_d=40$~MeV, the orbital and binding energy of the final state are changed according to \cite{chazono}. }
    \label{ratio}
\end{table}

\section{Summary  \label{conclusions}}

We have presented a numerical assessment for the first order perturbative correction to the ADWA description of ($d$,$p$) reactions that has been proposed in \cite{J14} for a class of separable rank-1 $n$-$p$ potentials. This correction arises due to additional nonlocal contribution to the optical potential to be used to calculate deuteron-target distorted waves. The corrected distorted waves were evaluated externally and   then used as input for the cross section calculations performed with
TWOFNR's help.
The first Weinberg component, representing exact CDCC calculations with (nonlocal) rank-1 Yamaguchi $n$-$p$ potential, was constructed from the CDCC solutions for deuteron-target scattering waves, which were obtained by running FRESCO with externally read-in coupling matrix elements in the Yamaguchi basis. FRESCO was also employed to read in this component and provide the CDCC cross sections.



Numerical calculations have shown that perturbative corrections affect mainly the partial waves with $\ell \leq \kappa R_N$ while higher $\ell$'s remain unaffected so that the overall influence of this correction depend on the relative contribution of lower and higher partial waves to the $(d,p)$ amplitude.
The first order perturbative correction has been applied to  $^{10}$Be($d$,$p$), $^{40,48}$Ca($d$,$p$) and $^{55}$Ni($d$,$p$) reactions for several beam energies large enough to neglect contributions from closed channels.
Overall, our results indicate corrections ranging from $2-25\%$ at forward angles  to up to $40-50$\% at scattering angles higher than $20$~degrees. The ratios of the ADWA+p.c. cross sections to ADWA and CDCC, shown in table \ref{ratio}, give an idea of how much spectroscopic factors and asymptotic normalization coefficients would change if perturbative calculation  replaced the ADWA or CDCC in the analysis of experimental $(d,p)$ data. These  ratios have been evaluated at 0~deg (1st peak), and at the first maximum of each differential cross section individually (2nd peak). The perturbative correction works better at higher angles, and at forward angles it sometimes brings the cross section closer to the CDCC predictions. 

In general, the first-order perturbative corrections as proposed in \cite{J14}, are not sufficient to bring the $(d,p)$ ADWA cross sections into agreement with exact three-body results. However, we should note that the approach of Ref. \cite{J14} is based on several assumptions, related to Green's functions properties, whose validity has not been thoroughly investigated. Further work should explore extension of the first-order perturbation theory beyond these assumptions. The interest for pursuing in this direction is motivated by difficulties in generalizing the CDCC approach to include nonlocal nucleon-target optical potentials, as well as clarifying the role of induced three-body effects in $(d,p)$ reactions beyond ADWA. First-order perturbation theory could advance our knowledge of these not yet understood physical problems.

\ack
This work has received funding from the  United Kingdom Science and Technology Facilities Council (STFC) under Grant No.  ST/P005314/1. LM is grateful to M.\ G\'omez-Ramos, 
D.\ Y.\ Pang,
and J.~Rangel for their help while learning FRESCO during lockdown, to P.~Descouvemont for his prompt assistance on R-matrix program, and to Y.~Chazono for useful clarifications. 


\appendix
\section*{Appendix}
\section{Scattering states for the Yamaguchi rank-1 separable potential \cite{yamaguchi}. \label{appendix}}

Neutron-proton scattering states $\mid \chi_{\ve{k}}\ra$ in the potential $V$ are defined as the limit $\epsilon \rightarrow 0+$ of the states
   
\beq \mid \chi^{(\epsilon)}_{\ve{k}}\ra=(2\pi)^{3/2}\frac{\imath \epsilon}{(E_k+\imath \epsilon-T_r-V)}\mid \ve{k}\ra,\label{cont}\eeq
where
\beq T_r=\frac{\ve{p}^2}{\mu_d},\,\,\,\la\ve{r}\mid \ve{k}\ra=\frac{\exp(\imath \ve{r}\cdot\ve{k})}{(2\pi)^{3/2}},\,\,E_k=\frac{\hbar^2k^2}{\mu_d}.
 \label{TrkE}\eeq
 and $\mu_d$ is the neutron-proton reduced mass. 
 The state $\mid \chi^{(\epsilon)}_{\ve{k}}\ra$ also satisfies
 \beq \mid \chi^{(\epsilon)}_{\ve{k}}\ra=(2\pi)^{3/2}\mid\ve{k}\ra+\frac{1}{(E_k+\imath\epsilon-T_r)}V\mid \chi^{(\epsilon)}_{\ve{k}}\ra,\label{cont2}\eeq
For a rank-1 separable  potential
\beq V=-\mid f\ra\la f\mid \label{V},\eeq 
 (\ref{cont2}) becomes 
\beq \mid \chi^{(\epsilon)}_{\ve{k}}\ra=(2\pi)^{3/2}\mid \ve{k}\ra-\frac{1}{(E_k+\imath\epsilon-T_r)}\mid f\ra\la f \mid \chi^{(\epsilon)}_{\ve{k}}\ra,\label{cont3}\eeq
and hence
\beq  \la f\mid\chi^{(\epsilon)}_{\ve{k}}\ra=(2\pi)^{3/2}\frac{\la f\mid\ve{k}\ra}{ 
(1+\la f\mid \frac{1}{(E_k+\imath \epsilon-T_r)}\mid f\ra)} .\label{cont4}\eeq
Substituting (\ref{cont4}) into (\ref{cont3}) gives the explicit expression for $\mid \chi^{(\epsilon)}_{\ve{k}}\ra $
\beq \mid \chi^{(\epsilon)}_{\ve{k}}\ra=(2\pi)^{3/2}\left(\mid \ve{k}\ra-\frac{1}{(E_k+\imath\epsilon-T_r)}\times\frac{\mid f\ra\la f\mid\ve{k}\ra}{ 
(1+\la f\mid \frac{1}{(E_k+\imath\epsilon-T_r)}\mid f\ra)}\right),\label{cont5}\eeq
For the Yamaguchi potential
\beq  \la \ve{r}\mid f \ra=N_1\frac{\exp(-\beta r)}{r},\label{rf}\eeq
we find
\beq \la \ve{k}'\mid f \ra =\frac{4\pi N_1}{(2\pi)^{3/2}}\frac{1}{(\beta^2+k'^2)},\label{kf}\eeq
and
\beq \la f\mid \frac{1}{(E_k+\imath \epsilon-T_r)}\mid f\ra = \frac{2\pi \mu_d N_1^2}{ \hbar^2\beta} \frac{1}{(k+\imath \frac{\epsilon}{2k}+\imath \beta)^2}. \label{fG0f}\eeq
A bound state $\mid\psi_0\ra$ satisfies
\beq \mid \psi_0\ra=\frac{1}{(-\epsilon_d-T_r)}V\mid \psi_0\ra
=\frac{1}{(\epsilon_d+T_r)}\mid f\ra\la f\mid \psi_0\ra,\label{bound}\eeq 
hence if $\mid \psi_0\ra \neq \mid 0 \ra$, the constants $\alpha, \beta, N_1$ must satisfy
\beq 
\la f\mid\frac{1}{(\epsilon_d+T_r)}\mid f\ra=1.\label{bound2}\eeq
By putting $k=\imath \alpha $, $\epsilon_d=\frac{\hbar^2}{\mu_d}\alpha^2,$ in the result (\ref{fG0f}) the condition (\ref{bound2}) gives
 \beq \frac{2\pi \mu_d N_1^2}{ \hbar^2\beta} \frac{1}{(\alpha+ \beta)^2}=1.\label{bound3}\eeq
 Therefore if the potential $V$ does support a bound deuteron the formula (\ref{fG0f}) can be replaced by
 \beq \la f\mid \frac{1}{(E_k+\imath \epsilon-T_r)}\mid f\ra= \frac{(\beta+\alpha)^2}{(k+\imath \frac{\epsilon}{2k}+\imath \beta)^2}. \label{fG0f2}\eeq
Using these explicit formulae in (\ref{cont5}) we find 
 \beq \la\ve{r}\mid \chi^{(\epsilon)}_{\ve{k}}\ra&=&\exp(\imath \ve{r}\cdot\ve{k})+\frac{2 \beta(\alpha+\beta)^2}{(k-\imath \alpha)(k-\imath \beta)^2(k+\imath(2\beta+\alpha))}  
 \eol
 &\times&
 \left(\frac{\exp(\imath kr)}{r}- \frac{\exp(-\beta r)}{r}\right),\label{cont6}
 \eeq
note that according to (\ref{cont6}) the scattered wave has $\ell=0$ only. 
 The coefficient of the scattered wave is the scattering amplitude $f(k,\theta)$ which in this case is independent of $\theta$ and related to the s-wave phase shift $\delta_k$, all other phase shifts vanish, by
 \beq f=\frac{\exp(\imath \delta_k) \sin \delta_k}{k}, \label{f}\eeq
 or equivalently
 \beq k\cot \delta_k=\imath k+\frac{1}{f}=\frac{k^4+(\alpha^2+2\alpha\beta+3\beta^2)k^2-\alpha\beta^2(\alpha+2\beta)}{2 \beta(\alpha+\beta)^2}, \label{cot}\eeq
 according to (\ref{cont6}). 
 The scattering length $a$ and the effective range $r_0$ are defined by the expansion in powers of $k$
 \beq k\cot\delta_k&&=-\frac{1}{a}+\frac{1}{2}r_0k^2+\dots.\label{coteff}\eeq
 From (\ref{cot}) we deduce
 \beq a&&=\frac{2 (\alpha+\beta)^2}{\alpha\beta(\alpha+2\beta)}\eol
 r_0&&=\frac{(\alpha^2+2\alpha\beta+3\beta^2}{ \beta(\alpha+\beta)^2}.\label{ar0}\eeq
The $\ell=0$ component of the scattering wavefunction (\ref{cont6}) can be written
 \beq \chi_0^{(+)}(k,r)&&=\frac{\sin(kr)}{kr}+\frac{\exp(\imath \delta_k) \sin \delta_k}{k}\left(\frac{\exp(\imath kr)}{r}- \frac{\exp(-\beta r)}{r}\right)\eol
 &&=\frac{\exp(\imath \delta_k)}{kr}(\sin(kr+\delta_k)-\sin \delta_k \exp(-\beta r)).\label{cont7}\eeq

\section*{References}

\begin{thebibliography}{99}

\bibitem{B1950}
Butler S T 1950
{\it Phys. Rev.} \href{https://doi.org/10.1103/PhysRev.80.1095.2}{{\bf 80} 1095--1096}

\bibitem{BKHN52}
Bhatia A, Huang K, Huby R, and Newns H 1952
{\it The London, Edinburgh, and Dublin
  Philosophical Magazine and Journal of Science} \href{https://doi.org/10.1080/14786440508520204}{{\bf 43} 485--500}

\bibitem{J14}
Johnson R C 2014
{\it J. Phys. G: Nucl. Part. Phys.}
\href{https://doi.org/10.1088/0954-3899/41/9/094005}{{\bf 41} 094005}

\bibitem{TIMOFEYUK2020103738}
Timofeyuk N and Johnson R C 2020
{\it Progr. Part. Nucl. Phys.}
\href{https://doi.org/10.1016/j.ppnp.2019.103738}{{\bf 111} 103738}

\bibitem{watanabe}
Watanabe S 1958
{\it Nucl. Phys.}
\href{https://doi.org/10.1016/0029-5582(58)90180-9}{{\bf 8} 484--492}

\bibitem{JS1970}
Johnson R C and Soper P J R 1970
{\it Phys. Rev. C}
\href{https://doi.org/10.1103/PhysRevC.1.976}{{\bf 1} 976--990}

\bibitem{JT74}
Johnson R C and Tandy P 1974
{\it Nucl. Phys. A}
\href{https://doi.org/10.1016/0375-9474(74)90178-X}{{\bf 235} 56--74}

\bibitem{RAW1975}
Rawitscher G H 1975
{\it Nucl. Phys. A}
\href{https://doi.org/10.1016/0375-9474(75)90393-0}{{\bf 241} 365--385}

\bibitem{AIKKRY87}
Austern N, Iseri Y, Kamimura M, Kawai M, Rawitscher G H and Yahiro M 1987
{\it Phys. Rep.}
\href{https://doi.org/10.1016/0370-1573(87)90094-9}{{\bf 154} 125--204}
  
\bibitem{Fad} Faddeev L D
1961
{\it Z. Eksp. Teor. Fiz.} 
\textbf{39} 1459;  1961 \textit{Sov. Phys. J. Exp. Theor. Phys.} \textbf{12} 1014

\bibitem{Fad96}
Austern N, Kawai M and Yahiro M 1996 
{\it Phys. Rev. C}
\href{https://doi.org/10.1103/PhysRevC.53.314}{{\bf 53} 314--321}

\bibitem{ND2011}
Nunes F M and Deltuva  A 2011
{\it Phys. Rev. C}
\href{https://doi.org/10.1103/PhysRevC.84.034607}{{\bf 84} 034607}

\bibitem{UDN2012}
Upadhyay N J , Deltuva A and Nunes F M 2012 {\it Phys. Rev. C}
\href{https://doi.org/10.1103/PhysRevC.85.054621}{{\bf 85} 054621}

\bibitem{chazono}
Chazono Y, Yoshida K and Ogata K 2017 
{\it Phys. Rev. C}
\href{https://doi.org/10.1103/PhysRevC.95.064608}{{\bf 95} 064608}
  
  \bibitem{Joh14} 
  Johnson R C and Timofeyuk N K 2014
  {\it Phys. Rev. C}
 \href{https://doi.org/10.1103/PhysRevC.89.024605}{{\bf 89} 024605}
  
\bibitem{Din19} 
Dinmore M J, Timofeyuk N K, Al-Khalili J S and Johnson R C 2019
 {\it Phys. Rev. C}
 \href{https://doi.org/10.1103/PhysRevC.99.064612}{{\bf 99} 064612}

\bibitem{weinberg63}
Weinberg S 1963
{\it Phys. Rev.}
\href{https://doi.org/10.1103/PhysRev.131.440}{{\bf 131} 440--460}

\bibitem{weinberg64}
Weinberg S 1964
{\it Phys. Rev.}
\href{https://doi.org/10.1103/PhysRev.133.B232}{{\bf 133} B232--B256}

\bibitem{PTJT13}
Pang D Y, Timofeyuk N K, Johnson R C and Tostevin J A 2013
 {\it Phys. Rev. C}
 \href{https://doi.org/10.1103/PhysRevC.87.064613}{{\bf 87} 064613}

\bibitem{yamaguchi}
Yamaguchi Y 1954
{\it Phys. Rev.}
\href{https://doi.org/10.1103/PhysRev.95.1628}{{\bf 95} 1628--1634}

\bibitem{HKB04}
Holt J D, Kuo T T S and Brown G E 2004
  {\it Phys. Rev. C}
  \href{https://doi.org/10.1103/PhysRevC.69.034329}{{\bf 69} 034329}

\bibitem{JBF08}
Jurgenson E D, Bogner S K, Furnstahl R J and Perry R J 2008
{\it Phys. Rev. C}
\href{https://doi.org/10.1103/PhysRevC.78.014003}{{\bf 78} 014003}

\bibitem{GT18}
G\'omez-Ramos M and Timofeyuk N K 2018
 {\it Phys. Rev. C}
 \href{https://doi.org/10.1103/PhysRevC.98.011601}{{\bf 98} 011601(R)}

\bibitem{Dd18}
Deltuva A 2018
 {\it Phys. Rev. C}
 \href{https://doi.org/10.1103/PhysRevC.98.021603}{{\bf 98} 021603(R)}

\bibitem{hulthen1}
Hulth\'en L and Laurikainen K V 1951 
 {\it Rev. Mod. Phys.}
 \href{https://doi.org/10.1103/RevModPhys.23.1}{{\bf 23} 1--9}

\bibitem{hulthen2}
Hulth\'en L and Nagel B C H 1953
  {\it Phys. Rev.}
  \href{https://doi.org/10.1103/PhysRev.90.62}{{\bf 90} 62--69}

\bibitem{Dgauss}
Yahiro M, Iseri Y, Kamimura M and Nakano M 1984
{\it Phys. Lett. B}
\href{https://doi.org/10.1016/0370-2693(84)90549-5}{{\bf 141} 19--22}

\bibitem{TWOFNR} Tostevin J A  University of Surrey corrected and updated version of the code \href{http://nucleartheory.eps.surrey.ac.uk/NPG/code.htm}{{\sc twofnr}} (of Toyama M, Igarashi M and Kishida N) and code {\sc front},  

\bibitem{satchler-DNR}
Satchler G 1983 {\it Direct Nuclear Reactions} (Oxford: Oxford University Press)

\bibitem{TJ13}
Timofeyuk N K and Johnson R C 2013
{\it Phys. Rev. Lett.}
\href{https://doi.org/10.1103/PhysRevLett.110.112501}{{\bf 110} 112501}

\bibitem{BTT2016}
Bailey G W, Timofeyuk N K and  Tostevin J A 2016
{\it Phys. Rev. Lett.}
\href{https://doi.org/10.1103/PhysRevLett.117.162502}{{\bf 117} 162502}

\bibitem{AS70}
Abramowitz M and Stegun I A 1970 {\it Handbook of Mathematical Functions} (New York: Dover) 

\bibitem{M09}
Michel N 2009
{\em Eur. Phys. J. A}
\href{https://doi.org/10.1140/epja/i2008-10738-7}{{\bf 42} 523}

\bibitem{Rmatrix}
Descouvemont P 2016
{\it Comp. Phys. Comm.}
\href{https://doi.org/10.1016/j.cpc.2015.10.015}{{\bf 200} 199--219}
  .

\bibitem{VTMLC91}
Varner P R, Thompson W, McAbee T, Ludwig E and Clegg T 1991
{\it Phys. Rep.}
\href{https://doi.org/10.1016/0370-1573(91)90039-O}{{\bf 201} 57}

\bibitem{KD02}
Koning A and Delaroche J 2003
{\it Nucl. Phys. A}
\href{https://doi.org/10.1016/S0375-9474(02)01321-0}{{\bf 713} 231--310}

\bibitem{FRESCO}
 Thompson I J 1988 {\it Comput. Phys. Rep.}
 \href{https://doi.org/10.1016/0167-7977(88)90005-6}{{\bf 7} 167}


\end{thebibliography}

\end{document}